\documentclass[twocolumn,showpacs,preprintnumbers,amsmath,amssymb,prb,aps]{revtex4-1}

\usepackage{epsfig}
\usepackage{graphicx}
\usepackage{dcolumn}
\usepackage{bm}

\begin{document}

\title{Orbital ordering in the geometrically frustrated MgV$_2$O$_4$: \emph{Ab initio}
electronic structure calculation}
\author{Sudhir K. Pandey}
\altaffiliation{Electronic mail: sk$_{_{-}}$iuc@rediffmail.com}
\affiliation{UGC-DAE Consortium for Scientific Research, University
Campus, Khandwa Road, Indore - 452001, India}

\date{\today}

\begin{abstract}
In the light of recent interesting experimental work on MgV$_2$O$_4$
we employ the density functional theory to investigate the crucial
role played by different interaction parameters in deciding its
electronic and magnetic properties. The strong Coulomb correlation
in presence of antiferromagnetic (AFM) coupling is responsible for
the insulating ground state. In the ground state the $d_{xz}$ and
$d_{yz}$ orbitals are ordered and intra-chain vanadium ions are
antiferromagnetically coupled. The calculation gives small
spin-orbit coupling (SOC), which provides a tilt of $\sim$11.3$^0$
to the magnetic moment from the $z$-axis. In the presence of weak
SOC and strong exchange coupling, the experimentally observed small
magnetic moment and low AFM transition temperature appear to arise
from spin fluctuation due to activeness of geometrical frustration.

\end{abstract}


\maketitle

\section{Introduction}
The orbital degree of freedom is an important entity in the
condensed matter physics which plays a crucial role in stabilizing
many exotic phases observed in the strongly correlated electron
systems.\cite{khomskii} When the degenerate $d$ orbitals of the
transition metals are partially filled then occupation of a
particular orbital at a particular site is expected to be dictated
by the occupation of another orbital at another site, which can lead
to various kind of orbital ordering (OO) similar to the spin
ordering. For example, in the case of LaMnO$_3$ degenerate $e_g$
orbitals are occupied by only one electron and predicted to show
antiferro-orbital ordering (AFOO) in the $xy$ plane in 1955 by
Goodenough.\cite{goodenough} The Coulomb correlation is found to
play an important role in the orbital physics of transition metal
oxides. However, it is still not clear whether it just enhances the
effect of lattice distortions or really drives the OO via
superexchange.\cite{pavariniKCF, pavariniLMO} In spite of the
ambiguity about the exact mechanism behind the OO, it is almost
clear that such OO is often accompanied by reduction in the crystal
symmetry. Thus in the geometrically frustrated system OO is expected
to relax the frustration leading to the formation of novel magnetic
phases earlier forbidden by the frustration.

Spinel vanadates with general formula AV$_2$O$_4$ (A-Cd, Mg, and Zn)
is an important geometrically frustrated system which has attracted
a great deal of attention for a decade because of OO induced
structural transition and formation of fascinating magnetic
phases.\cite{mamiya,nishiguchi,masashige,reehuis,tsunetsugu,tchernyshyov,lee,matteo,radaelli,maitra,wheeler}
All the studied compounds show cubic to tetragonal transition at low
temperature and paramagnetic (PM) to AFM transition at slightly
lower
temperature.\cite{mamiya,nishiguchi,masashige,reehuis,radaelli} The
consensus is emerging among the researchers regarding the OO induced
structural transition, however the exact pattern of OO remains a
matter of
controversy.\cite{tsunetsugu,tchernyshyov,matteo,maitra,wheeler}
Moreover, the exact role of SOC along with its strength in
stabilizing the magnetic and orbital ordering is yet to be decided
in this series of compounds.\cite{tsunetsugu,tchernyshyov,maitra} In
the present work we would like to address these issues for
MgV$_2$O$_4$ (MVO). The recent experimental work on this compound
has shown quite different results in comparison to the well studied
ZnV$_2$O$_4$ (ZVO).\cite{wheeler} In the tetragonal phase the space
group of MVO is \emph{I$\overline{4}$m2} whereas that of ZVO is
\emph{I4$_1$/amd}. The magnetic moment (MM) of V ion in MVO is
$\sim$0.47 $\mu_B$ which is $\sim$0.15 $\mu_B$ less than that in
ZVO. Such a reduced value of MM in ZVO is due to large but negative
contribution from the orbital part of MM. \cite{maitra} However,
experimental data of MVO do not suggest such a large contribution
from the orbital part  and indicating to a deeper reason for
observed small MM. Moreover, the MM is also seen to make an angle of
$\sim$8$^0$ with the $z$-axis indicating to a weak SOC in MVO
compound.

Here we explore the role played by spin and orbital degrees of
freedom in deciding the electronic and magnetic properties of MVO by
using \emph{ab initio} electronic structure calculations. The AFM
interaction in the presence of strong Coulomb correlation is found
to be crucial in driving the system to insulating ground state. The
$d_{xz}$ and $d_{yz}$ orbitals get ordered in the tetragonal phase
and OO becomes more robust in the presence of AFM interaction. The
spin and orbital part of MM is found to $\sim$1.4 and -0.2 $\mu_B$,
respectively. In the light of this result and in the presence of
large exchange coupling of $\sim$58 meV one can suggest that the
geometrical frustration may be responsible for the experimentally
observed low AFM transition temperature ($\sim$42 K) and small MM
($\sim$0.47 $\mu_B$).

\section{Computational details}
The non-magnetic (NM), ferromagnetic (FM), and AFM solutions of MVO
are obtained by using {\it state-of-the-art} full-potential
linearized augmented plane wave (FP-LAPW) method.\cite{elk} The
lattice parameters and atomic positions used in the calculations are
taken from the literature.\cite{wheeler} The muffin-tin sphere radii
automatically set in the calculations are 1.5664, 1.7045, and 1.4943
Bohr for Mg, V, and O atoms, respectively. For the exchange
correlation functional, we have adopted recently developed
generalized gradient approximation (GGA) form of Perdew {\em et
al.}\cite{perdew} The effect of on-site Coulomb interaction is also
considered within GGA+$U$ formulation of the density functional
theory.\cite{gga+u} In GGA+$U$ method the $U$ and $J$ are used as
parameters. We varied $U$ from 3-5 eV and fixed $J$=0.5 eV. We found
similar results for all values of $U$. The results correspond to
$U$=4 eV are only discussed in the manuscript. In order to see the role
of orbital degrees of freedom on the electronic and magnetic
properties of the compound SOC is also considered in the
calculations. The self-consistency was achieved by demanding the
convergence of the total energy to be smaller than 10$^{-4}$
Hartree/cell.

\section{Results and discussions}
The atomic arrangements in the unit cell are shown in Fig. 1. It is
evident from Fig. 1(a) that each V atom is surrounded by six O atoms
forming an octahedron. The octahedra are edge shared to each other.
The small trigonal distortion splits six V-O bonds of an octahedron
in two groups containing three bonds each with bond length of 2.016 and
2.033 {\AA}. The four nearby V atoms form a regular tetrahedron
with an edge of length 2.98 {\AA}. Each of the tetrahedron is
surrounded by four neighboring tetrahedra via corner sharing and
forming the chains of V atoms, see Fig. 1(b). In the tetragonal
phase the tetrahedra become distorted with the V-V bond length of
2.971 and 2.980 {\AA}. Such a small distortion would reduce the
geometrical frustration and is expected to give rise to novel
electronic and magnetic phases.

In order to know the exact ground state of the compound we obtained
various magnetic solutions using tetragonal structure.  The total
density of states (TDOS) correspond to these solutions are shown in
Fig. 2. It is clear from Figs. 2(a)-(d) that the GGA solutions
provide metallic state as opposed experimentally observed insulating
behavior. This result is not surprising as the GGA underestimates
the Coulomb correlation among the 3$d$ electrons which is often
found to be responsible for insulating ground state of the
transition metal
oxides.\cite{anisimov,sudhirLCO,sudhirSNRO,sudhirSCO} The NM and FM
solutions obtained from GGA+$U$ calculations also provide metallic
state as evident from Figs. 2(e) and (f). This indicates that there
may be a deeper reason for the insulating ground state of the
system. At this juncture, it is important to note that the FM
solution of ZVO within LSDA+$U$ is found to derive insulating ground
state of the compound.\cite{maitra} In order to know the exact cause
for insulating ground state of MVO we performed AFM calculations.
The AFM solution creates a soft gap and in presence of SOC it
provides a hard gap of $\sim$0.16 eV, see Figs. 2(g) and (h). It is
important to note that the increased value of $U$ enhances the band
gap in both the cases. Thus present work clearly establishes that
the AFM coupling of V moments in presence of strong on-site Coulomb
interaction is responsible for insulating ground state and SOC
provides robustness to the insulating property of the compound.

Now we discuss the effect of different interaction parameters on the
electronic structure of the compound. First we start with GGA
results. In the absence of magnetic interaction there are large
density of states (DOS) at the Fermi level ($\epsilon_F$) with
dominating contribution from the V 3$d$ states. This may be
considered as a signature of magnetic ground state under Stoner
theory. The magnetic interaction reduces the DOS at $\epsilon_F$ by
$\sim$50\% due to exchange splitting of the bands contributing at
the $\epsilon_F$ and providing almost half-metallic state, see Fig.
2(b). Moreover, the energy of FM solution is found to be $\sim$0.65
eV/fu (fu=formula unit) less than that of NM solution indicating the
magnetic ground state. The energy difference between the band edge
of the up and down spins may be considered as a measure of exchange
interaction which is found to be $\sim$0.4 eV. The MM of V is found
to be $\sim$1.25 $\mu_B$. Interestingly, FM interaction induces
finite MM ($\sim$0.14 $\mu_B$) at the Mg ions occupying 2$c$
(0,1/2,1/4) Wyckoff sites. The total MM/fu comes out to be $\sim$3.6
$\mu_B$, which corresponds to $S$$\approx$1 state of the V$^{3+}$
ion. It is evident from Fig. 2(c) that the AFM interaction among the
V moments reduces the DOS at the $\epsilon_F$ drastically ($\sim$4
times less than that of FM). The $\epsilon_F$ lies at the minima of
DOS which is a reminiscence of the pseudo-gap. The AFM interaction
reduces the band width (BW) of the system and the BW of the deeper
bands decreases by $\sim$0.3 eV. Further, it decreases the MM of V
by $\sim$0.1 $\mu_B$ and does not create any MM at Mg sites. The
energy/fu of the AFM solution is $\sim$0.16 eV less than that of FM
solution. This is a clear evidence of the AFM ground state whose
spin ordering will be discussed in the later part of the manuscript.
The inclusion of SOC at this stage does not have any significant
effect on the electronic structure of the compound as evident from
Fig. 2(d). By comparing the energy of the AFM and AFM+SOC solutions
one can get the rough estimate of SOC strength of V 3$d$ electron as
the contribution from Mg and O atoms is expected to be negligibly
small. Our GGA calculation gives the strength of SOC of $\sim$5 meV
for the V 3$d$ electrons.

Figs. 2(e)-(h) depict the effect of on-site Coulomb interaction
among the V 3$d$ electrons on the electronic properties of the
compound in the presence of various interaction parameters. On-site
Coulomb correlation reduces the BW as it localizes the electrons. In
the case of NM and FM solutions there is a drastic decrease in the V
3$d$ DOS at the $\epsilon_F$ due to transfer of spectral weight
(earlier contributing at $\epsilon_F$) away from it. Moreover, FM
interaction gives rise to a perfectly half-metallic state with a
band gap of $\sim$2.5 eV in the down-spin channel. In the presence
of AFM interaction among the V spins the system becomes insulating
due to formation of upper and lower Hubbard bands. Further,
inclusion of SOC increases the separation between upper and lower
Hubbard bands. The exchange interaction estimated from the FM
solution is $\sim$0.5 eV which is 0.1 eV more than that obtained
from simple GGA calculation. This enhancement is attributed to
increased Hund's coupling strength due to increased spatial
localization of V 3$d$ electrons, which also enhances the MM at V
sites by $\sim$0.2 $\mu_B$. The total MM/fu comes out to be 4
$\mu_B$, which corresponds to $S$=1 state of the V$^{3+}$ ion. Such
a large value of magnetic moment at V site is in sharp contradiction
with the experimentally observed small magnetic
moment.\cite{wheeler} This clearly indicates that some other
parameters are playing important role in deciding the magnetic
properties of the compound. On comparing the energy of various
solutions we find that the AFM state is the true ground state of the
system as energy of NM$>$FM$>$AFM.

In order to study the role of orbital degrees of freedom we have
performed FM GGA+$U$ calculations in both cubic and tetragonal
phases. In the cubic phase the occupancies of $d_{xz}$ and $d_{yz}$
orbitals are same at every V sites, whereas for the tetragonal phase
the occupancies of these orbitals are found to be different at
different site which is a direct evidence of OO taking place in the
tetragonal structure. Moreover, the OO pattern does not depend on
the nature of magnetic interaction as evident from Table 1 where we
have listed the occupancy of $d_{x^2-y^2}$,\cite{comment1} $d_{xz}$,
and $d_{yz}$ orbitals of four V atoms forming the tetrahedron and
obtained from FM and AFM solutions. It is evident from the table
that each site is occupied by $d_{x^2-y^2}$ orbital. The V1 and V2
sites are mainly occupied by $d_{xz}$ orbital and that of V3 and V4
sites by $d_{yz}$. At this juncture it is important to note that OO
is observed in the PM phase of the spinel vanadates and in the PM
phase there is a local MM at the V site and hence magnetic solutions
would provide the better representation of the PM state in
comparison to NM solution. Table 1 also indicates the AFM coupling
between V1 and V2 (V3 and V4) and FM coupling between V1 and V4 (V2
and V3). Interestingly, AFM interaction appears to provide more
stability to the OO as the occupancy of $d_{yz}$ ($d_{xz}$) orbital
at V1 and V2 (V3 and V4) sites is found to decrease by $\sim$0.08.
Moreover the energy/fu of AFM solution is also $\sim$0.24 eV less
than that of FM solution suggesting the AFM ground state. This
energy difference between AFM and FM solutions is $\sim$80 meV less
than that obtained from GGA solution. This highlights the importance
of Coulomb correlation in establishing the AFM ground state. The MM
of vanadium ions corresponds to FM and AFM solutions are found to be
$\sim$1.43 and 1.35 $\mu_B$, respectively.

As mentioned in the introduction that the OO is normally considered
as a cause for the structural transition in the PM phase of spinel
vanadates. However, based on present work it is difficult to say
whether OO is the cause of structural transition or it is just an
effect of it. In order to understand the cause of OO seen in
different transition metal oxides mainly two mechanism exist in the
literature which are purely electronic and structural in
origin.\cite{khomskii,pavariniKCF, pavariniLMO} To separate out
these two contributions to the OO, Pavarini \emph{et al.} have
carried out beautiful work on two canonical OO systems viz. KCuF$_3$
and LaMnO$_3$, where they have used LDA+DMFT
method.\cite{pavariniKCF, pavariniLMO} It is important to note that
GGA+$U$ method used in the present work is a static mean-field
theory whereas LDA+DMFT used in the work of Pavarini \emph{et al.}
is a dynamical mean-field theory and hence better in approximation.
Thus in order to know the exact cause of OO in the MVO compound work
in line with Pavarini \emph{et al.} is desirable.

Now we study the effect of SOC on the magnetic state of the
compound. The GGA+$U$+SOC solutions also give AFM ground state as
the energy/fu of AFM solution is found to be $\sim$0.18 eV less than
that of FM solution. As mention above the AFM interaction provides
more stability to the OO which further enhances the orbital moment
as evident from Table 2 where we have shown the spin ($S$), orbital
($L$) and total ($J$) moment of the V ion corresponds to FM and AFM
solutions. The orbital part of MM in the AFM state comes out to be
$\sim$-0.2 $\mu_B$ which is $\sim$7 times less than the spin part of
MM suggesting the weak SOC in MVO with respect to ZVO\cite{maitra}
where large orbital moment of -0.75 $\mu_B$ has been reported. The
direction of total MM is found to be $\sim$11.3$^0$ away from the
$z$-axis. The small value of orbital moment and the direction of
total magnetic moment are in consonance with the experimental
findings where neutron scattering studies have revealed the small
orbital moment and MM is tilted at $\sim$8$^0$ from the
$z$-axis.\cite{wheeler} However, the above calculated values of spin
and orbital moments cannot account for the experimentally observed
magnetic moment of 0.47 $\mu_B$. At this point it is important to
note that the experimentally estimated magnetic moment of 0.63
$\mu_B$ for ZVO is well accounted by taking into account the
calculated large but negative orbital moment of -0.75
$\mu_B$.\cite{maitra} This indicates that the orbital sector of the
MVO is not as influential as found in the ZVO in deciding the
magnetic state of the V ions. Thus there may be a deeper reason for
the experimentally observed small magnetic moment in the MVO which
will be discussed in the next paragraph. The final spin and orbital
ordering patterns obtained from the calculations are shown in Fig.
3. The spins are forming AFM chains in the $x$ and $y$ directions
and nearest neighbor AFM chains are connected by lines with FM
ordering. The AFM chains are accompanied by ferro-orbital ordering
(FOO) where $d_{xz}$ and $d_{yz}$ orbitals are occupied along the
$x$ and $y$ directions, respectively. The AFOO supports the
formation of FM chains where neighboring sites are alternatively
occupied by $d_{xz}$ and $d_{yz}$ orbitals. These spin and orbital
ordering patterns are in accordance with the Goodenough-Kanamori
schemes. Here it is important to note that historically Goodenough
has given the semi-covalent scheme for explaining the experimentally
observed complex magnetic structures in La$_{1-x}$Ca$_x$MnO$_3$ and
predicted different OO corresponds to different spin
arrangements.\cite{goodenough} According to this scheme length of FM
bond should be greater than that of AFM bond. However, we have
observed opposite behavior as AFM bond is 0.03 {\AA} larger than the
FM bonds.

As mentioned above that the calculated small orbital moment of -0.2
$\mu_B$ for MVO cannot account for the experimentally observed value
of total MM $\sim$0.47 $\mu_B$, whereas calculated large orbital
moment of -0.75 $\mu_B$ for ZVO provides a good description of its
experimentally observed total MM of 0.63 $\mu_B$. These results
appear to suggest that the geometrical frustration is still active
in the tetragonal phase of MVO which can give rise to spin
fluctuation at low temperature. Such spin fluctuation is expected to
reduce the MM drastically. The level of frustration in magnetic
systems is defined by frustration index $f$ $\equiv$
$|$$\theta$$_{CW}$$|$/$T^*$, where $\theta$$_{CW}$ is the
Curie-Weiss temperature and $T^*$ is the critical temperature at
which the system ultimately develops long-range spin
order.\cite{gardner} Higher the value of $f$ more will be the level
of frustration. Thus the above conjecture about the activeness of
frustration can further be tested by estimating the Heisenberg
exchange interaction strength ($J_H$) between V moments on which
$\theta$$_{CW}$ depends. The rough estimate of it can be found by
mapping the energies of FM and AFM solutions to the Heisenberg
Hamiltonian.\cite{sudhirEPL} Our calculation gives $J_H$$\approx$58
meV. Using this value of $J_H$ we have estimated the AFM transition
temperature based on mean-field theory and it comes out to be
$\sim$925 K, which is closer to the experimental value of
$\theta$$_{CW}$.\cite{mamiyaSSC} Using experimental $T^*$$\approx$42
K we have estimated the value of $f$$\approx$22. Such a large value
of $f$ further strengthens the conjecture about the activeness of
frustration in MVO compound. Here it is important to note that the
GGA+$U$ formulation of density functional theory is a mean-field
theory, which is not capable of addressing issue related with spin
fluctuations directly. Thus, to address this one needs to go beyond
the mean-field theory.

\section{Conclusions}
In conclusion, we have investigated the electronic and magnetic
properties of a geometrically frustrated MgV$_2$O$_4$ by using
\emph{ab initio} electronic structure calculations. This compound is
a Mott-insulator and its insulating ground state is arising due to
combined effect of strong Coulomb correlation and AFM interaction.
The $d_{xz}$ and $d_{yz}$ orbitals are found to be ordered in the
tetragonal phase. The spins are forming AFM ordered chains along the
$x$ and $y$ directions and making an angle $\sim$11.3$^0$ with the
$z$-axis. The SOC is weak and geometrical frustration appears to be
active in deciding the magnetic state of the system.





FIG. 1. (Color online) Atomic arrangement of the unit cell. The
formation of VO6 octahedra and V4 tetrahedra are shown in (a) and
(b), respectively.

FIG. 2. (Color online) Evolution of total density of states (TDOS)
with various interaction parameters. Please see the text for the
details.

FIG. 3. (Color online) (a) Intra-chain antiferromagnetic ordering
along the $x$ and $y$ directions. (b) spin and orbital arrangements
at the tetrahedron level.


Table 1: Occupancies of $d_{x^2-y^2}$, $d_{xz}$, and $d_{yz}$
orbitals and magnetic moments of four V atoms (viz. V1, V2, V3, and
V4 forming the tetrahedron, see Fig. 3(b)) corresponding to FM and
AFM (in brackets) solutions obtained from GGA+$U$ ($U$=4 eV)
calculations.

\vspace{2ex}
\begin{tabular}{|c|c|c|c|c|}
  \hline
   & V1 & V2 & V3 & V4 \\
     \hline
  $d_{x^2-y^2}$ & 0.62(0.65) & 0.62(0.65) & 62(0.65) & 62(0.65) \\
    \hline
  $d_{xz}$ & 0.64(0.66) & 0.64(0.66) & 0.18(0.1) & 0.18(0.1) \\
    \hline
  $d_{yz}$ & 0.18(0.1) & 0.18(0.1) & 0.64(0.66) & 0.64(0.66) \\
    \hline
  MM ($\mu_B$) & 1.43(1.35) & 1.43(-1.35) & 1.43(-1.35) & 1.43(1.35) \\
  \hline
\end{tabular}

Table 2: The expectation value $x$, $y$ and $z$ components of spin
($S$), orbital ($L$) and total ($J$) moment of V ion obtained from
FM and AFM (in brackets) GGA+$U$+SOC ($U$=4 eV) solutions.

\begin{tabular}{|c|c|c|c|}
  \hline
    & $x$ & $y$ & $z$ \\
    \hline
  $S$ & $\sim$0($\sim$0) & $\sim$0($\sim$0) & 0.71(0.67) \\
  \hline
  $L$ & -0.02(-0.1) & $\sim$0($\sim$0) & -0.07(-0.17)\\
  \hline
  $J$ & -0.02(-0.1) & $\sim$0($\sim$0)& 0.64(0.5) \\
  \hline

\end{tabular}



\begin{thebibliography}{99}
\bibitem{comment1} $d_{x^2-y^2}$ orbial in the body-centered tetragonal coordinate system
is same as the $d_{xy}$ orbital in the face-centered cubic
coordinate system.
\end{thebibliography}
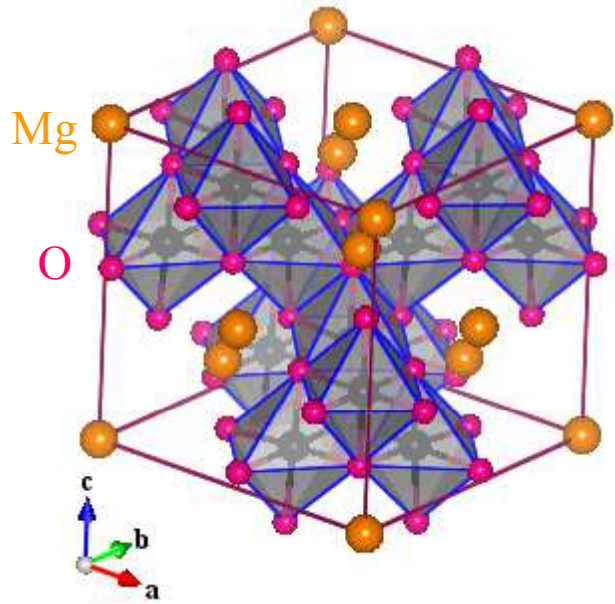
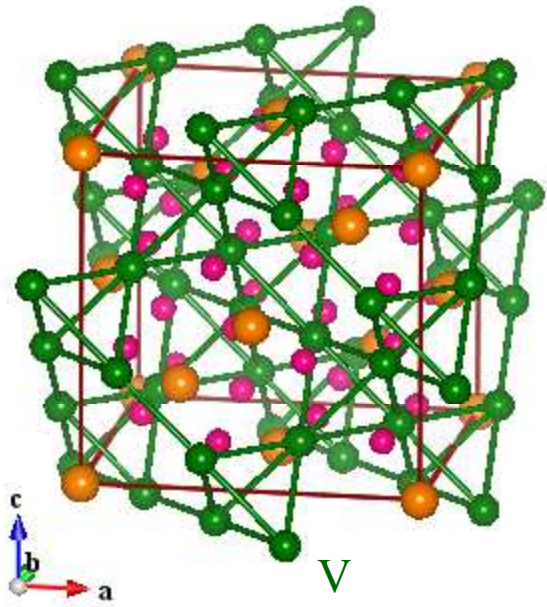
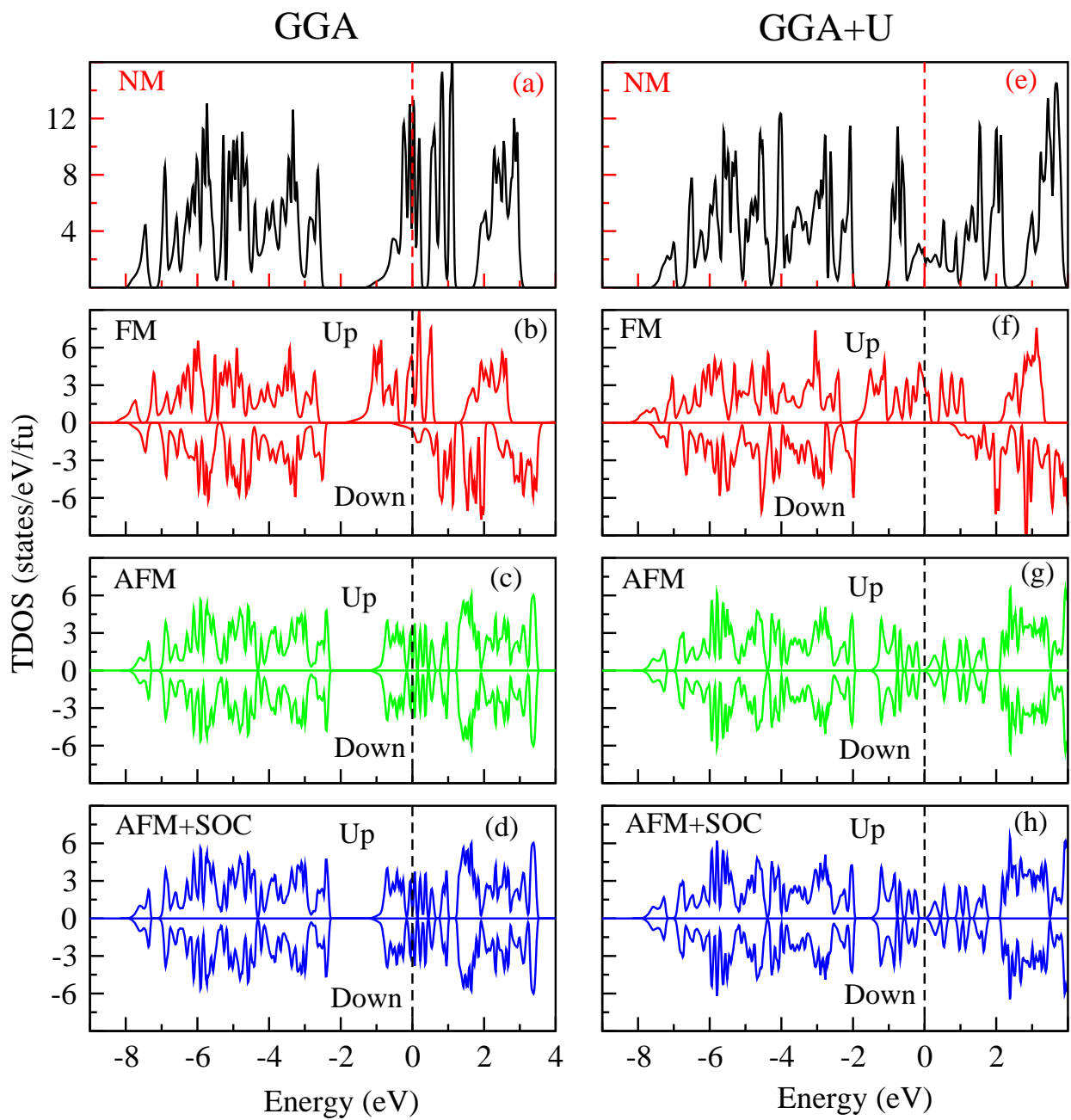
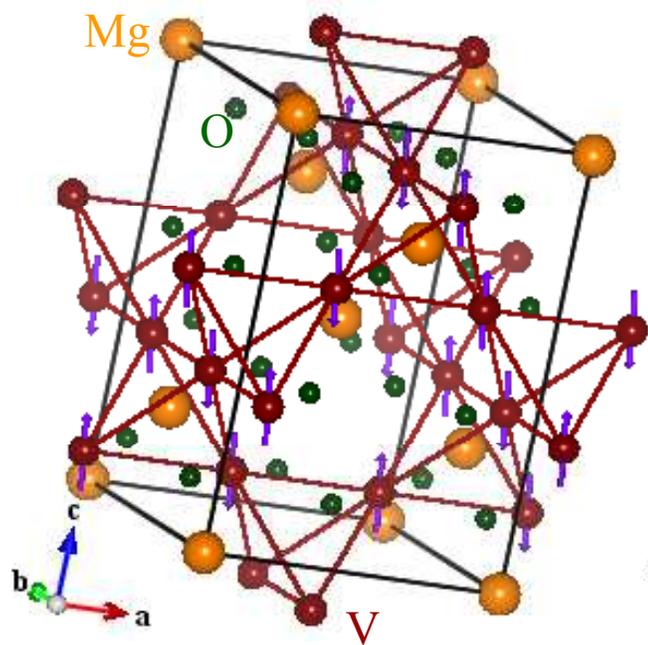
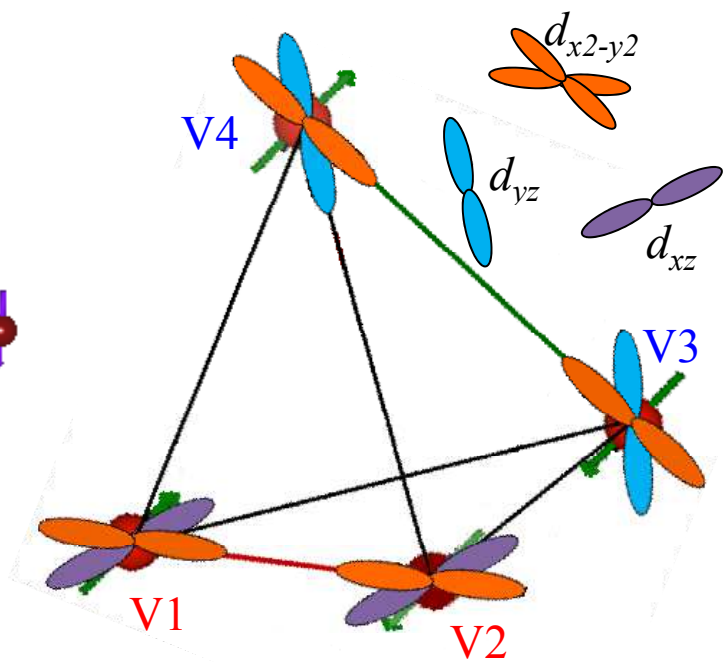
\end{document}